\documentclass[dvips]{acta}
\usepackage{supertabular,lscape,epsfig}
\usepackage{amssymb}
\usepackage{amsmath}
\usepackage[T1]{fontenc}

\SetPages{0}{0}

\SetVol{76}{2026}

\usepackage{lmodern}

\newcommand{\RES}{\Re(\sigma)}
\newcommand{\IMS}{\Im(\sigma)}

\newcommand{\EKIN}{\mathcal{E}_{\rm kin}}
\newcommand{\PEX}{\mathcal{P}_{\rm ex}}

\newcommand{\PC}{\mathcal{P}_{\rm c}}
\newcommand{\PJ}{\mathcal{P}_{\rm J}}
\newcommand{\PPV}{\mathcal{P}_{\rm pV}}

\newcommand{\PBND}{\mathcal{P}_{\rm bnd}}

\newcommand{\NMC}{N_{\rm c}}
\newcommand{\NMH}{N_{\rm ha}}
\newcommand{\NMS}{N_{\rm sg}}
\newcommand{\NMI}{N_{\rm in}}

\newcommand{\CHIE}{\chi_{\rm eff}}
\newcommand{\LTEF}{\log(T_{\rm eff})}

\newcommand{\SN}{S_{\rm N}}

\newcommand{\mFig}[1]{Fig.~\ref{fig:#1}}
\newcommand{\mEq}[1]{Eq.~(\ref{eq:#1})}

\DeclareMathOperator{\sgn}{sgn}

\begin{document}

\begin{Titlepage}

\Title{Mode power and stability in radial post-AGB pulsations}

\Author{Zalewski, J.}
{Independent researcher \\
e-mail: jan.zalewski.a2@gmail.com}

\end{Titlepage}

\Abstract{

We examine the relation between mode power and excitation rate for radial pulsations along a sequence of high-luminosity post-AGB envelope models. The net thermodynamic exchange power, including boundary contributions, is compared with the excitation rate and the inertia, compression and horizontal area deformation norms entering the balance relation. We find that positive net exchange power can accompany damping, while negative net exchange power can accompany excitation. Along an individual mode sequence, the power may reverse sign repeatedly without a change of stability. These reversals are associated with changes in the sign of the effective norm. Including the compression contribution in the pressure-force power does not remove this behaviour and can extend the intervals of sign disagreement. Thus, although spatial exchange integrals identify positive and negative contributions to the modal power, the sign of the global power alone does not generally determine mode stability.

}
{stars: oscillations, stars: AGB and post-AGB, instabilities}

\section{Introduction}

Work integrals are widely used in nonadiabatic pulsation studies to evaluate mode growth and damping rates and to identify pulsational instability domains on the HR diagram (Unno et al. 1989, Dziembowski 1994, Pamyatnykh 1999). Their application to strongly nonadiabatic pulsations was discussed by Glatzel (1994). Saio, Georgy and Meynet (2013) employed such a relation in their analysis of blue supergiants and used cumulative and differential work integrals to examine the driving and damping of ordinary and strange modes. 

The spatial distribution of work is also used to identify the stellar layers contributing to pulsation driving and damping. Recent examples include the linear calculations discussed by Reese (2022) and the nonlinear early post-AGB calculations of Gautschy (2023), in which pressure--volume work over individual pulsation cycles was evaluated throughout the envelope. Daszy\'nska-Daszkiewicz et al. (2025) used work integral information as well as an instability parameter $\eta$ based on work integral in the discussion of unstable modes.

In our previous papers (Zalewski 2026f and references therein) we have obtained the balance relation for nonadiabatic linear pulsations and found it contains several terms that are substantial for the balance to hold. The resulting balance relation is of the form $\PEX\sim -\gamma(\NMI+\NMC+\NMH+\NMS)$ and thus depends on the inertia, compression, horizontal area deformation and gravitational-stratification norms. Thus a full nonadiabatic balance relation contains not only the inertia term proportional to $\EKIN$ but several others, which modify the relation between mode power and excitation rate.

The norm terms contribute with different signs and their sum can lead to a change of sign of the effective norm. Because of that the relation between the thermodynamic exchange power including the surface terms and the kinetic energy of pulsation may enter an anomalous regime where the sign relations between the thermodynamic power and mode excitation rate are reversed (Zalewski 2026e). The higher overtone radial modes for post-AGB models are examples of damped anomalous modes occurring for $\LTEF<4$. Anomalous excited low frequency modes also occur at higher temperatures (Zalewski 2026ef).

In this paper we will examine the behavior of the power generated by the mode and its excitation rate by analyzing radial nonadiabatic pulsation modes for a range of $\LTEF$ covering most of the post-AGB range for high luminosity models.

\section{The balance relation}
In Zalewski (2026cd) we have derived the balance relation for nonadiabatic radial and nonradial pulsation. This relation is of the form
\begin{equation}
	\PEX+\PBND=2\gamma\EKIN\CHIE,
\label{eq:BalanceEq}
\end{equation}
where $\PEX$ denotes the real part of the complex thermodynamic exchange power, dependent on $p \dot{\overline{s}}$, defined in Zalewski (2026f), and $\PBND$ is the power due to boundary terms. The excitation rate $\gamma=\RES$. For perturbations with time dependence $\exp(\omega t)$, the dimensional frequency is related to $\sigma$ as $\omega=\Omega_0\sigma$ (Zalewski 2026c). $\EKIN$ is the mode's kinetic energy. The $\CHIE$ is the effective norm factor term (Zalewski 2026e) given below. 

The mode kinetic energy may be expressed in terms of mode inertia norm ($\NMI$) as 
\[
\EKIN=-\NMI/2,
\]
and the effective norm $\CHIE$ for radial pulsation is given by
\[
\CHIE=1+\frac{\NMC+\NMH}{\NMI}.
\]
The two norm terms are negative $\NMI<0$, $\NMC<0$ while $\NMH$ for nonradial pulsations can become positive or negative depending on the eigenfunctions of radial and horizontal displacements, but for radial pulsation it is positive. Thus if this term becomes sufficiently positive for a particular eigenfrequency the $\CHIE<0$ and the sign of the net exchange power and mode's excitation rate reverses leading to anomalous modes (Zalewski 2026ef).

The boundary contribution is represented by $\PBND$. Depending on assumed boundary conditions (Zalewski 2026a) this term may become substantial compared to $\PEX$. In our calculations we select boundary conditions such as to make this term small (see however Zalewski 2026e).

The forms for these terms (Zalewski 2026f) for radial pulsation are given by
\begin{equation}
  \begin{aligned}
  	\PEX&=-\int_{x_b}^{x_s} CA_7\Re(p\,\overline{\sigma s})\,dx, \\
  	\PBND&=-\left[C\Re(p\,\overline{\sigma d})\right]^{x_s}_{x_b}, \\
  	\NMI&=-\int_{x_b}^{x_s}C A_3 A_2 |\sigma|^2 |d|^2\, dx, \\
  	\NMC&=-\int_{x_b}^{x_s}C A_4 |p|^2\, dx, \\
  	\NMH&=4\int_{x_b}^{x_s} C A_3 |d|^2\, dx,
  \end{aligned}
\label{eq:BalanceInt}  
\end{equation}
where the independent variable is $x=\ln(r/R_{\odot})$ and $x_b$ and $x_s$ are the inner and outer boundaries of the pulsation region. The overline symbol denotes complex conjugation. The other symbols in the above equations are given in Zalewski (2026c).

From \mEq{BalanceInt} it is seen that the thermodynamic exchange power depends on pressure ($p$) and entropy ($s$) perturbations, while the other terms depend on radial displacement ($d$) or pressure perturbations.

With this notation the balance relation for radial pulsation is written as
\begin{equation}
	\mathcal{P}=\PEX+\PBND=-\gamma(\NMI+\NMC+\NMH).
\label{eq:RadialBalanceEq}
\end{equation}

Sometimes, instead of the thermodynamic exchange power, the pressure-volumne power $\mathcal{P}_{\rm pV}$ is used to compute the work integral
\[
\PPV=\PEX+\PC=\Re\int_{x_b}^{x_s}Cp\overline{\sigma q}\,dx,
\]
where $q$ is the relative Lagrangian volume perturbation and $\PC=\gamma \NMC$.

The integrated pressure-force power $\PJ$ is then given by
\[
\PJ=\PPV+\PBND.
\]

Since work expressions based on both $\PEX$ and $\PPV$ are used in the literature, it is also of interest to examine whether including the compression contribution changes this anomalous sign relation.

For anomalous modes $\CHIE<0$, which for radial pulsation implies that 
\[
\NMH>|\NMI+\NMC|.
\]

Combining the above definitions with \mEq{RadialBalanceEq} gives
\[
\begin{aligned}
\PJ&=\PEX+\PBND+\PC \\
   &=-\gamma(\NMI+\NMH) \\
   &=\gamma\left(|\NMI|-\NMH\right).
\end{aligned}
\]
Consequently, both the net exchange power $\mathcal{P}=\PEX+\PBND$ and the integrated pressure-force power $\PJ$ have signs opposite to $\gamma$. For excited anomalous modes both powers are negative, whereas for damped anomalous modes both are positive. Thus inclusion of the compression contribution does not change the anomalous relation between power and excitation rate.

Even for modes with $\CHIE>0$, the sign of $\PJ$ is opposite to that of $\gamma$ when $|\NMI|<\NMH<|\NMI+\NMC|$, so the sign reversal extends beyond the anomalous regime defined by the net exchange power $\PEX+\PBND$.

With the adopted definitions
\begin{equation}
\mathcal{P}=2\gamma\EKIN\CHIE, \quad \CHIE=1+\frac{\NMC+\NMH}{\NMI},
\label{eq:PCHIEq}
\end{equation}
whereas
\begin{equation}
\PJ=2\gamma\EKIN\chi_{\rm J}, \quad \chi_{\rm J}=1+\frac{\NMH}{\NMI}.
\label{eq:PJEq}
\end{equation}
Therefore 
\[
\CHIE-\chi_{\rm J}=\frac{\NMC}{\NMI}\geq 0,
\]
and thus
\[
\chi_{\rm J}\leq\CHIE.
\]
It thus follows that when $\CHIE<0$ then $\chi_{\rm J}<0$, but there may be modes for which $\chi_{\rm J}<0$ while $\CHIE>0$ thus the range of anomalous modes based on the sign difference between $\PJ$ and $\gamma$ is broader than that when the sign of $\mathcal{P}$ is compared to the sign of $\gamma$. This means that if $\PJ$ is used to predict instability regions on the HR diagram, broader instability ranges may be inferred than when computing power terms based on the thermodynamic exchange term. However, as will be discussed in the following section, using the sign of either power without checking the norm components entering the balance relation may lead to inferred instability intervals that differ substantially from those determined by the sign of the excitation rate.

\section{Comparison of mode power and excitation rate}

\subsection{Analysed modes}

\begin{figure}[htb]
	\includegraphics{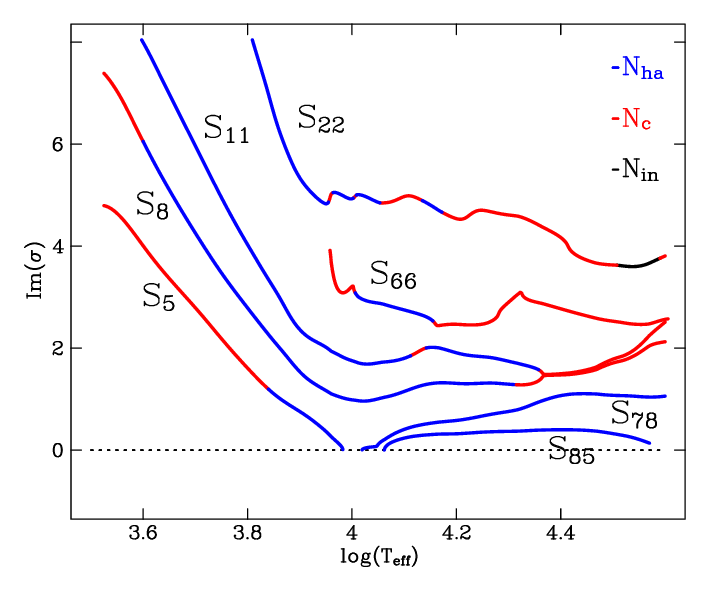}
	\FigCap{The dependence of the oscillation frequency $\IMS$, on effective temperature for the sequences of modes described in text. The curves are color-coded as - blue in regions where the horizontal area term is largest, red where the compression term has largest magnitude and black where the inertia term has largest magnitude. Modes are labelled by mode sequence numbers. }
	\label{fig:Fig1}
\end{figure} 

We select representative mode sequences from our survey of radial pulsations of high luminosity, $L/M=14500 L_{\odot}/M_{\odot}$, post-AGB envelopes. At high $L/M$ ratios, as may be seen from Fig.~1 in Zalewski (2026e) or from a similar figure in Gautschy (1993), mode sequences are often subject to avoided crossings (Gautschy and Glatzel 1990). 
The selected sequences of modes are shown in \mFig{Fig1}. The color coding is the same as in Zalewski (2026f). 

Among the selected sequences, $\mathsf{S}_{22}$ can be traced throughout $3.55\leq\LTEF\leq 4.6$. At $\LTEF=3.8$, it corresponds to the sixth radial overtone, with $\IMS=8.37$. The  other sequences considered are $\mathsf{S}_5$ which is followed from $\LTEF=3.525$, where it corresponds to the fundamental mode and continues with $\IMS$ decreasing to zero near $\LTEF=3.978$ and two sequences - viz. a 1-ov ($\mathsf{S}_8$) and 2-ov ($\mathsf{S}_{11}$) also followed from the same effective temperature and continuing to $\LTEF=4.6$ with an avoided crossing near $\LTEF=4.365$. 

Two additional sequences shown in \mFig{Fig1} are the low frequency, high temperature anomalous modes starting near $\LTEF=4.05$ and followed to $\LTEF=4.6$ - viz. $\mathsf{S}_{85}$ and $\mathsf{S}_{78}$. The low frequency anomalous mode $\mathsf{S}_{85}$ which becomes excited near $\LTEF=4.33$ was described in Zalewski (2026ef). These modes, as well as the low temperature ($\LTEF<4$) strange modes visible in Fig.~1 in Zalewski (2026e) are not analyzed here.

From \mFig{Fig1} it may be seen that the modes $\mathsf{S}_{8}$ and $\mathsf{S}_{11}$ undergo avoided crossing near which they become compression term dominated, while the sequence $\mathsf{S}_{22}$ proceeds from horizontal area dominated for $\LTEF<4$ to compression dominated below $\LTEF=4.5$ followed by regions where the inertial term has larger magnitude than the remaining two.

The mode visible as sequence $\mathsf{S}_{66}$ undergoes several avoided crossings with other modes in the range of effective temperatures $4<\LTEF<4.3$ producing rapid bends in its projection onto the $(\LTEF,\IMS)$ plane. 

\subsection{Scaling of quantities}
For a given mode sequence the net exchange power, as well as norms span multiple orders of magnitude over the range of effective temperatures analyzed here corresponding to the blue-ward post-AGB crossing of the HR diagram. Thus plots of such sequences versus effective temperature would be dominated by the regions with largest values, not necessarily the regions which matter for the stability of the mode. Therefore we have introduced a normalization enabling comparisons of norms on the same plot for the whole range of effective temperatures by utilizing 
\[
S_{\rm N} = -\NMI-\NMC>0
\]
as the normalization factor (both norms are negative). In this way the scaled norm ratios are defined as
\[
S_{\rm in}=\frac{-\NMI}{\SN}, \qquad S_{\rm c}=\frac{-\NMC}{\SN}, \qquad S_{\rm ha}=\frac{\NMH}{\SN}.
\]
The norm ratios show relative contributions and are independent of the eigenfunction normalization. The sum $S_{\rm in}+S_{\rm c}=1$.

The balance relation using these ratios may be written as
\[
\PEX+\PBND=-\gamma(\NMI+\NMC)(1-S_{\rm ha}),
\]
and the effective norm $\CHIE$ as
\[
\CHIE=\frac{1-S_{\rm ha}}{S_{\rm in}}.
\]

With this scaling, when the horizontal area deformation term exceeds the sum of the absolute values of other two  
\[
S_{\rm ha}>1 \iff \NMH>|\NMI+\NMC| \iff \CHIE<0,
\]
and the mode becomes an anomalous exchange one. Thus the condition for a mode to be anomalous ($\CHIE<0$) may be equivalently expressed as $S_{\rm ha}>1$ with the latter form exposing the role of the horizontal area deformation term.

The exchange power components may also be scaled in the same way as
\[
\frac{\PEX+\PBND}{\SN}=\gamma -\gamma\frac{\NMH}{\SN}, 
\]
and thus readily compared to $\gamma$. 

However since both the net exchange power terms as well as $\gamma$ change substantially over the covered range of effective temperatures a diagram using the exchange power and excitation rate would also be dominated by largest magnitude regions whereas the regions important for the present discussion are where the net exchange or excitation rate cross zero. For that purpose the exchange power as well as the excitation rate are scaled as
\[
S(A)=\sgn(A)\log\left(1+\frac{|A|}{S_0}\right),
\]
where $A$ is the quantity under consideration. For the present case these will be mainly $S_{\rm P}=S(\mathcal{P})$ and $S_\gamma$ as well as $S_{\rm J}=S(\PJ)$. The $S_0$ is arbitrary and was chosen as $S_0=10^{-4}$. With this scaling and for the arguments much larger than $S_0$ the scaling is approximately logarithmic, while for small values a near linear dependence on the quantity is obtained. This scaling function retains signs and zeros.

\subsection{Net exchange power and excitation rate}

\begin{figure}[htb]
	\includegraphics{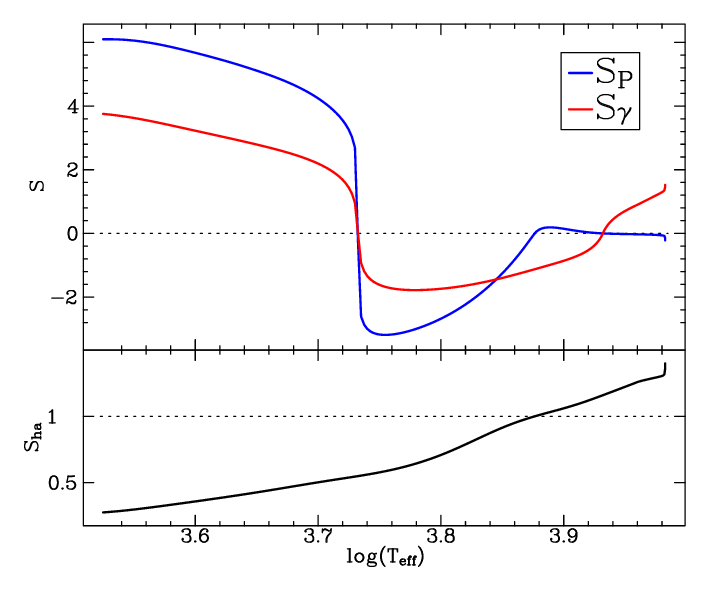}
	\FigCap{Upper panel - plot of $S_{\rm P}$ and $S_\gamma$ for the range of effecitve temperatures for the fundamental mode - sequence $\mathsf{S}_5$. While excitation rate $\gamma$ crosses zero two times - at $\LTEF=3.73$ and $3.93$ the net exchange power $\mathcal{P}$ crosses zero three times, at the same temperatures as $\gamma$ and at $\LTEF=3.88$ where $S_{\rm ha}=1$. Lower panel - plot of scaled horizontal area deformation norm.}
	\label{fig:Fig2}
\end{figure}

The net exchange power and excitation rate as well as the ratio $S_{\rm ha}$ are shown in \mFig{Fig2} for the fundamental mode for effective temperature range $3.52 \leq\LTEF< 3.98$. The mode is excited $\gamma>0$ for $\LTEF\leq 3.73$. This temperature may be considered as the blue edge of the radial fundamental mode instability strip for the analyzed envelope model. At higher temperatures the mode is damped ($\gamma<0$) to become excited again at $\LTEF\geq 3.93$. The ratio $S_{\rm ha}$ increases with effective temperature and exceeds unity near $\LTEF=3.88$ and the mode becomes anomalous for higher effective temperatures, which leads for $\LTEF>3.93$ to $\gamma>0$ while $\mathcal{P}<0$. Thus for this mode the sign agreement between net exchange power ($\mathcal{P}$) and excitation rate holds until $S_{\rm ha}=1$ and reverses at higher temperatures.

The mode sequence $\mathsf{S}_{85}$ discussed in Zalewski (2026f) is another example of anomalous behavior of exchange power relative to excitation rate. For this mode the exchange power and excitation rate cross zero at the same effective temperature $\LTEF=4.33$ but are of opposite signs. Thus inferring the stability of this mode based on exchange power would lead to instability regions opposite to those that the mode exhibits.

\begin{figure}[htb]
	\includegraphics{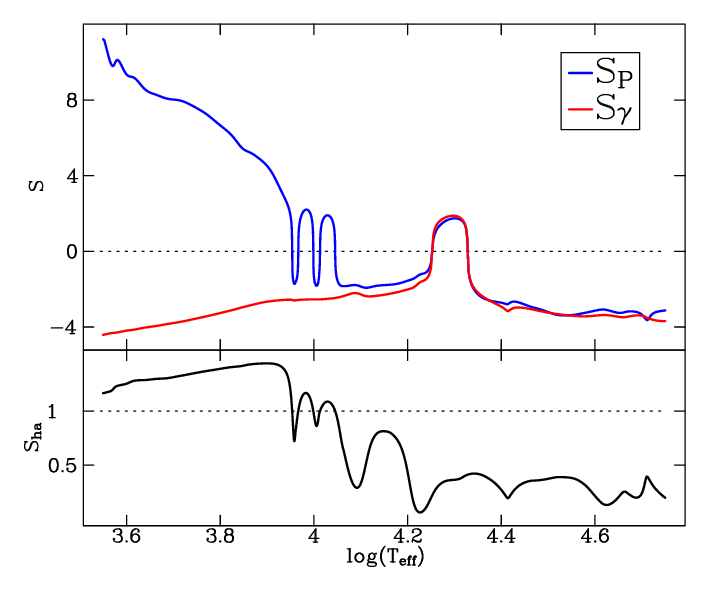}
	\FigCap{Same plot as \mFig{Fig2} but for the sixth overtone mode - sequence $\mathsf{S}_{22}$. The $\mathcal{P}$ crosses zero several times, while the excitation rate remains negative for $\LTEF<4.2$. At higher temperatures, where $S_{\rm ha}<1$ for this mode, it becomes one of normal relation between the signs of net exchange power and excitation rate. }
	\label{fig:Fig3}
\end{figure}

As may be seen from \mFig{Fig3} the sixth-overtone sequence $\mathsf{S}_{22}$ exhibits repeated reversals of net exchange power while the mode remains damped. The ratio $S_{\rm ha}$ exceeds unity at low effective temperatures and crosses unity several times within $3.94\leq\LTEF\leq4.05$. For higher temperatures it remains below unity becoming small. Each crossing of $S_{\rm ha}=1$ produces a sign reversal of $\mathcal{P}$ while the excitation rate remains negative. Provided that at the crossing of $S_{\rm ha}=1$ the excitation rate is nonzero the $\mathcal{P}$ has to cross zero. Thus the regions of positive net exchange power are determined by two factors: crossing of zero by excitation rate, and $S_{\rm ha}$ becoming unity. This makes the net exchange power not a diagnostic of instability regions but a quantity that is computed for the eigenmode and required to satisfy the balance relation. This does not preclude the use of the cumulative integral form for the net exchange to diagnose the regions contributing to mode driving (see eg. Zalewski 2026f).

\subsection{Effect of boundaries on the power-excitation rate relations}
\begin{figure}[htb]
	\includegraphics{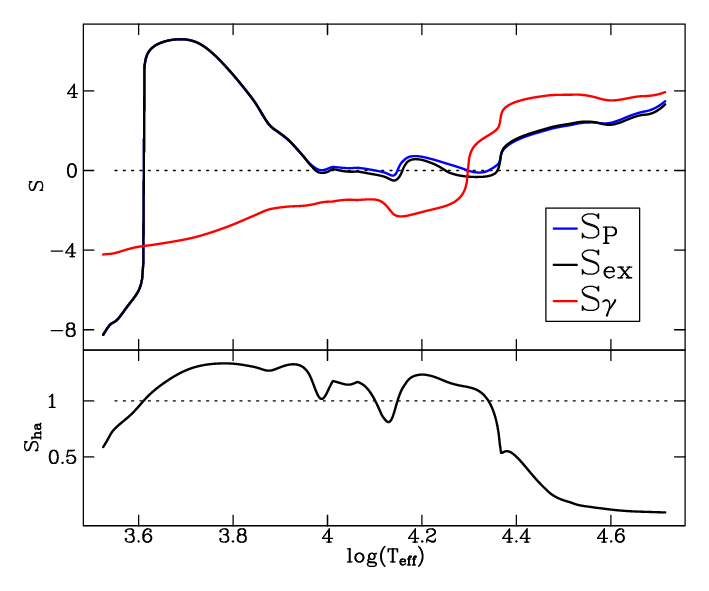}
	\FigCap{Same plot as \mFig{Fig2} but for the second overtone mode - sequence $\mathsf{S}_{11}$. The exchange power $\PEX$ crosses zero near $\LTEF=4.25$ while the net exchange power $\mathcal{P}$ at $\LTEF=4.33$ the difference is due to the contribution to the net power by the surface power $\PBND$ term. }
	\label{fig:Fig4}
\end{figure}

The role of surface terms ($\PBND$) on mode stability may be seen on \mFig{Fig4} for the second overtone mode - sequence $\mathsf{S}_{11}$. This mode crosses with the first overtone ($\mathsf{S}_{8}$) near $\LTEF=4.37$ but becomes unstable at $\LTEF=4.3$ where both $\gamma=\mathcal{P}=0$. The mode is anomalous in the range $3.61\leq\LTEF\leq 4.34$ with a narrow range where $S_{\rm ha}<1$ near $\LTEF\approx 4.1$. The scaled exchange power $S_{\rm ex}=S(\PEX)$ is also shown.

In case of this mode the difference between net power $\mathcal{P}$ and the exchange power $\PEX$ is not large, however since both powers are small, and close to zero, the boundary power term $\PBND$ may affect the balance. As may be seen from \mFig{Fig4}, the zero crossing of $\gamma$ coincides with that of $\mathcal{P}$, whereas $\PEX$ crosses zero at a different effective temperature because of the boundary contribution $\PBND$. 

For some other modes, the boundary contribution is substantial. For example, after a series of avoided crossings within $4.15<\LTEF<4.51$, the third-overtone sequence becomes strongly damped at higher effective temperatures. In this strongly damped regime, $\PEX$ differs substantially from $\mathcal{P}$ because of the boundary contribution. A similar effect was described for low-temperature strange modes in Zalewski (2026e).

\subsection{Pressure-force power and excitation rate}
\begin{figure}[htb]
	\includegraphics{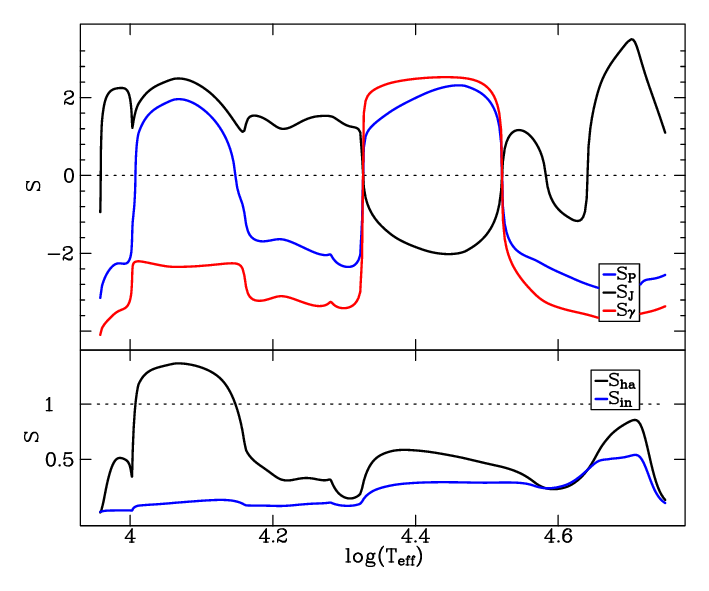}
	\FigCap{Upper panel - plot of $S_{\rm P}$ and $S_{\rm J}$ as well as $S_{\gamma}$ for the mode - sequence $\mathsf{S}_{66}$. Lower panel - scaled horizontal area deformation and inertia terms. It may be seen that the plots of $\mathcal{P}$ and $\PJ$ do not overlap. The two powers change sign at different effective temperatures and therefore identify different intervals of positive power. }
	\label{fig:Fig5}
\end{figure}

In \mFig{Fig5} the behavior of net exchange power $\mathcal{P}$ is compared to that of the pressure-force power $\PJ$. Since $\PJ$ is determined by \mEq{PJEq} a plot of $S_{\rm in}$ is shown alongside $S_{\rm ha}$. 

As may be seen from \mFig{Fig5} the pressure-force power crosses zero when either $\gamma=0$ or when $S_{\rm in}=S_{\rm ha}$ which follows from \mEq{PJEq} and the opposite signs of inertia and horizontal area deformation norms. Thus regions where $\PJ>0$ are determined by different conditions than those for $\mathcal{P}$. For this mode there are three zero crossings by $\PJ$ caused by $\chi_{\rm J}=0$, and two due to $\gamma=0$, while for $\mathcal{P}$ there are two crossings due to $\CHIE=0$ in addition to the crossings where excitation rate changes sign. Thus neither $\mathcal{P}$ nor $\PJ$ delimit the same regions of mode instability as obtained from the analysis of $\gamma(\LTEF)$. Thus use of any of these power terms may lead to anomalous modes, for which the sign of the power term differs from the sign of excitation rate.

\section{Conclusions}
We have examined the behavior of two forms of mode power quantities - viz. the net exchange power $\mathcal{P}=\PEX+\PBND$ and the pressure-force power $\PJ=\PPV+\PBND$ both of which follow from the work form $\mathcal{K}_{\rm s}=-p\dot{\overline{q}}$ and differ by the inclusion of compression response power term $\PC$ in $\PJ$, while for $\mathcal{P}$ based balance relation the compression term is included in the effective norm (see \mEq{RadialBalanceEq}).

The relation between mode power and its kinetic energy involves effective norm factor $\chi$ dependent on the form of adopted power expression. The two cases examined in this paper are given by \mEq{PCHIEq} and \mEq{PJEq}. From these forms of balance relation it follows that when the excitation rate $\gamma=0$ then the power term has to become zero. This is seen in the examples of modes analyzed in the previous sections. However, either effective norm factor can also vanish through cancellation of its constituent norms, making the corresponding power zero even when \(\gamma\ne0\). Thus the intervals of positive power need not coincide in position or number with the instability intervals determined from the excitation rate along a mode sequence.

Both effective norm factors may change sign due to the horizontal area deformation norm ($\NMH$) being of opposite sign (for radial pulsations) to the inertia ($\NMI$) and compression norms ($\NMC$). For the balance relation based on net exchange power $\mathcal{P}$ and effective norm factor $\CHIE$ the relevant quantity is $S_{\rm ha}$. When $S_{\rm ha}$ crosses unity while $\gamma\neq0$, the net exchange power changes sign without a change of mode stability. For the pressure-force power, the corresponding crossing occurs at $S_{\rm ha}=S_{\rm in}$.

We have established that $\chi_{\rm J}\leq\CHIE$, which means that every mode for which $\CHIE<0$ has also $\chi_{\rm J}<0$. These modes are anomalous with respect to both $\mathcal{P}$ and $\PJ$. There are also modes for which $\chi_{\rm J}<0<\CHIE$. These modes are anomalous with respect to $\PJ$ only. The anomalous sign relation between mode power and excitation rate may occur for both cases. The fact that for an excited mode the global power is negative does not mean that it is negative throughout the whole pulsation region.

The balance relation is evaluated on the computed eigenmode and the boundary power ($\PBND$) forms part of that balance. The boundary contribution power term should be included in the net power so that the power vanishes wherever the excitation rate vanishes.

We have analyzed balance relation for radial modes, but its formulation for nonradial ones given in Zalewski (2026d) contains the same norm terms as for radial pulsation plus an additional gravitation-stratification norm term $\NMS$ which, like $\NMH$ may be of opposite sign to the inertia and compression norms. Thus anomalous modes are possible also for nonradial pulsations.

As we have shown the sign of power alone is insufficient to determine mode stability. It must be interpreted together with the corresponding effective norm factor.

\end{document}